\documentclass[a4paper,12pt]{article}
\pdfoutput=1
\usepackage{bbm}
\usepackage{amsmath,times,amssymb,latexsym,graphics,braket,color,ascmac,subfigure}
\usepackage{amsmath}
\usepackage[dvipdfm]{graphicx}
\usepackage{bm}
\usepackage{hyperref}
\usepackage{cite}

\begin{document}
\begin{titlepage}
\begin{flushright}
{\small \today}
 \\
\end{flushright}

\begin{center}

\vspace{1cm}

\hspace{3mm}{\bf \Huge Notes on Pseudo Entropy Amplification} \\[3pt] 

\vspace{1cm}

\renewcommand\thefootnote{\mbox{$\fnsymbol{footnote}$}}
Yutaka {Ishiyama}${}^{1}$\footnote{chyt22001@g.nihon-u.ac.jp}, Riku {Kojima}${}^{1,2}$\footnote{kojima@rcnp.osaka-u.ac.jp}, Sho Matsui${}^{1}$\footnote{chho18051@g.nihon-u.ac.jp}, and Kotaro {Tamaoka}${}^{1}$\footnote{tamaoka.kotaro@nihon-u.ac.jp}

\vspace{5mm}

${}^{1}${\small \sl Department of Physics, College of Humanities and Sciences, Nihon University, \\Sakura-josui, Tokyo 156-8550, Japan}\\
${}^{2}${\small \sl Research Center for Nuclear Physics (RCNP), Osaka University, \\Ibaraki 567-0048, Japan.}\\

\end{center}

\vspace{5mm}

\noindent
\abstract
We study pseudo entropy for a particular linear combination of entangled states in qubit systems, two-dimensional free conformal field theories (CFT), and two-dimensional holographic CFT. We observe phenomena that the pseudo entropy can be parametrically large compared with the logarithm of the dimension of Hilbert space. We call these phenomena pseudo entropy amplification. The pseudo entropy amplification is analogous to the amplification of the weak value. In particular, our result suggests the holographic CFT does not lead the amplification as long as the non-perturbative effects are negligible. We also give a heuristic argument when such (non-)amplification can occur.

\end{titlepage}
\setcounter{footnote}{0}
\renewcommand\thefootnote{\mbox{\arabic{footnote}}}
\tableofcontents
\flushbottom
\section{Introduction and Summary}
Quantum entanglement is a key concept to understanding various aspects of physics, including resources of quantum computation~\cite{nielsen2002quantum}, critical phenomena~\cite{Vidal:2002rm,Calabrese:2004eu}, topological phases~\cite{Kitaev:2005dm,Levin:2006zz}, and the emergence of spacetime in holography~\cite{Ryu:2006bv,Hubeny:2007xt,Lewkowycz:2013nqa,Maldacena:2001kr,Swingle:2009bg,VanRaamsdonk:2009ar,VanRaamsdonk:2010pw,Pastawski:2015qua}. The entanglement entropy, von Neumann entropy of the reduced density matrix, is one of the most fundamental measures to quantify the quantum entanglement for the bipartite pure state. 

Recently, two states generalization of the entanglement entropy has been introduced, called pseudo entropy~\cite{Nakata:2020luh}. 
Let two states $\ket{\psi},\ket{\varphi}$ belong to a bipartite Hilbert space $\mathcal{H}_A\otimes\mathcal{H}_{B}$. Pseudo entropy is defined as von Neumann entropy associated with reduced \textit{transition} matrix,
\begin{align}
    S(\tau^{\psi|\varphi}_A)=-\mathrm{Tr}_{A}\left(\tau^{\psi|\varphi}_A\log\tau^{\psi|\varphi}_A\right), 
\end{align}
where the (reduced) transition matrix is given by
\begin{align}
    \tau^{\psi|\varphi}_A&=\mathrm{Tr}_{B}\,\tau^{\psi|\varphi},\\
    \tau^{\psi|\varphi}&=\dfrac{|\psi\rangle\langle\varphi|}{\langle\varphi|\psi\rangle}.
\end{align}
If the two states are the same, $\psi=\varphi$, the pseudo entropy reduces to the entanglement entropy for subsystem $A$. Importantly, we assume the two states have non-trivial overlap, $\langle\varphi|\psi\rangle\neq0$. The pseudo entropy has been found to be a natural generalization of holographic entanglement entropy~\cite{Ryu:2006bv,Hubeny:2007xt,Lewkowycz:2013nqa}, which is dual to area of the minimal surface in the Euclidean time-dependent backgrounds. 

Since the transition matrix is non-Hermitian, pseudo entropy can take complex values in general. Nevertheless, we have a nice quantum information theoretic interpretation for the real part: the real part of pseudo entropy can be identified with the number of distillable EPR pairs when averaging over intermediate states for fixed initial and final states. (Refer to \cite{Nakata:2020luh, Akal:2021dqt} for more details.) There are several applications of pseudo entropy in literature. One can use the pseudo entropy as a diagnosis of whether two states belong to the same quantum phase or not~\cite{Mollabashi:2020yie, Mollabashi:2021xsd, Nishioka:2021cxe}. The entanglement entropy for non-Hermitian systems~\cite{Chang:2019jcj} can be understood as a concrete example of pseudo entropy as the left and right energy eigenstates are different in non-Hermitian Hamiltonians. Other interesting applications include the characterization of entanglement Hamiltonian~\cite{Goto:2021kln} and gravitational setup with the absence of the time reflection symmetry~\cite{Miyaji:2021lcq, Akal:2021dqt}. Refer also to further developments relevant to the pseudo entropy~\cite{Camilo:2021dtt, Murciano:2021dga, Mukherjee:2022jac,Guo:2022sfl}.

It is well-known that von Neumann entropy is bounded above by the logarithm of the dimension of the Hilbert space $\log\dim\mathcal{H}$. On the other hand, pseudo entropy $S(\tau^{\psi|\varphi}_A)$ can be larger than $\log\dim\mathcal{H}_A$. We will call this property {\it amplification} of pseudo entropy. Note that this amplification is also relevant to the classification of quantum phases~\cite{Mollabashi:2020yie, Mollabashi:2021xsd, Nishioka:2021cxe} mentioned above. 

In this paper, we study such amplification by focusing on a typical situation where two states $\ket{\psi},\ket{\varphi}$ have a small but non-zero overlap. 
Similar amplification can occur for the quantity called weak value~\cite{Aharonov:1988xu,dressel2014colloquium},
\begin{align}
    \mathrm{Tr}(\tau^{\psi|\varphi}\mathcal{O})=\dfrac{\langle\varphi|\mathcal{O}|\psi\rangle}{\langle\varphi|\psi\rangle},
\end{align}
where the transition matrices naturally appear. 
The weak value of $\mathcal{O}$ can also take a larger value than the maximum eigenvalue of $\mathcal{O}$. However, this is not always the case even if $\braket{\varphi|\psi}$ is small. 
We study the similar setup in qubit systems (section \ref{sec:qubit}), two-dimensional free CFT (section \ref{sec:free}) and two-dimensional holographic CFT (section \ref{sec:hcft}). Notably, while we observe the amplification in qubit systems and free CFT, we do not see it in holographic CFT as long as non-perturbative contributions are negligible. 

The organization of the rest of this paper is as follows. In section \ref{sec:qubit}, we study pseudo entropy amplification in qubit systems and discuss heuristic criteria for when pseudo entropy can (and cannot) be amplified. In section \ref{sec:free}, we study amplification in two-dimensional free CFT. In section \ref{sec:hcft}, we study (non-)amplification in two-dimensional holographic CFT. In section \ref{sec:con}, we conclude and discuss future prospects.

{\bf Note added:} While we were finalizing this paper, we became aware of \cite{Guo:2022sfl} where the authors study the time-evolution of pseudo entropy by quench with the linear combination of operators in 2D CFTs. Although our motivation and parameter regime of interest are different from their paper, there are overlaps with our section \ref{sec:free}. 

\if(
\section{Review of pseudo entropy}
In this section, we briefly review some properties of pseudo entropy \cite{Nakata:2020luh}. Let us assume our Hilbert space is bipartite, $\mathcal{H}_A\otimes\mathcal{H}_{B}$. We define pseudo entropy as von Neumann entropy associated with reduced \textit{transition} matrix,
\begin{align}
    S(\tau^{\psi|\varphi}_A)=-\mathrm{Tr}_{A}\,\tau^{\psi|\varphi}_A\log\tau^{\psi|\varphi}_A, 
\end{align}
where the (reduced) transition matrix is defined as
\begin{align}
    \tau^{\psi|\varphi}_A&=\mathrm{Tr}_{B}\,\tau^{\psi|\varphi},\\
    \tau^{\psi|\varphi}&=\dfrac{|\psi\rangle\langle\varphi|}{\langle\varphi|\psi\rangle}.
\end{align}
If two states are the same, $\psi=\varphi$, the pseudo entropy reduces to the entanglement entropy for subsystem $A$. Therefore, we can see the pseudo entropy as a two-state generalization of entanglement entropy. Importantly, we assume two states have non-trivial overlap, $\langle\varphi|\psi\rangle\neq0$.
Transition matrix is no longer Hermitian in general, so not guaranteed to be diagonalizable. Even in such cases, we can still define trace of such matrices. As is the entanglement entropy, we define the pseudo R\'{e}nyi entropy,
\begin{align}
    S^{(n)}(\tau^{\psi|\varphi}_A)=\dfrac{\log\mathrm{Tr}_{A}\,(\tau^{\psi|\varphi}_A)^n}{1-n}. \label{eq:pre}
\end{align}
Thus, strictly speaking, we have to first define this R\'{e}nyi version and then define the pseudo entropy as $n\rightarrow1$ limit of \eqref{eq:pre}. 

We note some basic properties of pseudo entropy. As like the entanglement entropy, if either $|\psi\rangle$ or $|\varphi\rangle$ is a product state, pseudo entropy becomes trivial,
\begin{align}
    S^{(n)}(\tau^{\psi|\varphi}_A)=0 \;\;\;(\textrm{if $|\psi\rangle$ or $|\varphi\rangle$ is a product state}).
\end{align}
We also have a mutuality,
\begin{align}
    S^{(n)}(\tau^{\psi|\varphi}_A)=S^{(n)}(\tau^{\psi|\varphi}_B).
\end{align}

Since the transition matrix is non-Hermitian, pseudo entropy can take complex value in general. Nevertheless, we have a nice quantum information theoretic interpretation for the real part: the real part of pseudo entropy can be identified with the number of distillable EPR pairs when averaging over intermediate states for fixed initial and final states. (Refer to \cite{Nakata:2020luh, Akal:2021dqt} for more details.)
)\fi

\section{Amplification in Qubit Systems}\label{sec:qubit}
In this section, we first discuss qubit systems as an warm up exercise.  Let us consider Bell states (EPR pairs),
\begin{align}
    \ket{\Phi^{\pm}}&=\dfrac{1}{\sqrt{2}}(\ket{0_A0_B}\pm\ket{1_A1_B}), \label{eq:psi}\\
    \ket{\Psi^{\pm}}&=\dfrac{1}{\sqrt{2}}(\ket{0_A1_B}\pm\ket{1_A0_B}). \label{eq:varphi}
\end{align}
In particular, these states are orthogonal to each other. In what follows, we will utilize $\ket{\Phi^{\pm}}$. We can repeat the same discussion for any different choice of two of four Bell states. The important point here is that these two states are both entangled states and orthogonal to each other. We cannot define transition matrix between these two states as we have no probability of transition from one state to another. Nevertheless, we can perturb one of two Bell states such that
\begin{align}
    \ket{\psi}&=\ket{\Phi^+}+\epsilon\ket{\Phi^-}, \\
    \ket{\varphi}&=\ket{\Phi^-},
\end{align}
then we have small but non-zero overlap, $\langle\varphi|\psi\rangle=\epsilon$.
Namely, we will consider a formal rare process $\psi\rightarrow\varphi$ that potentially amplifies the weak value as we make the parameter $\epsilon$ smaller. As we will see soon, it is also the case for pseudo entropy. Although this setting may seem somewhat artificial, we choose it because it is convenient for comparison with calculations in field theories.

\begin{figure}[t]
    \centering
    \subfigure[$\textrm{Im}(\epsilon)=0$]{\resizebox{80mm}{!}{\includegraphics{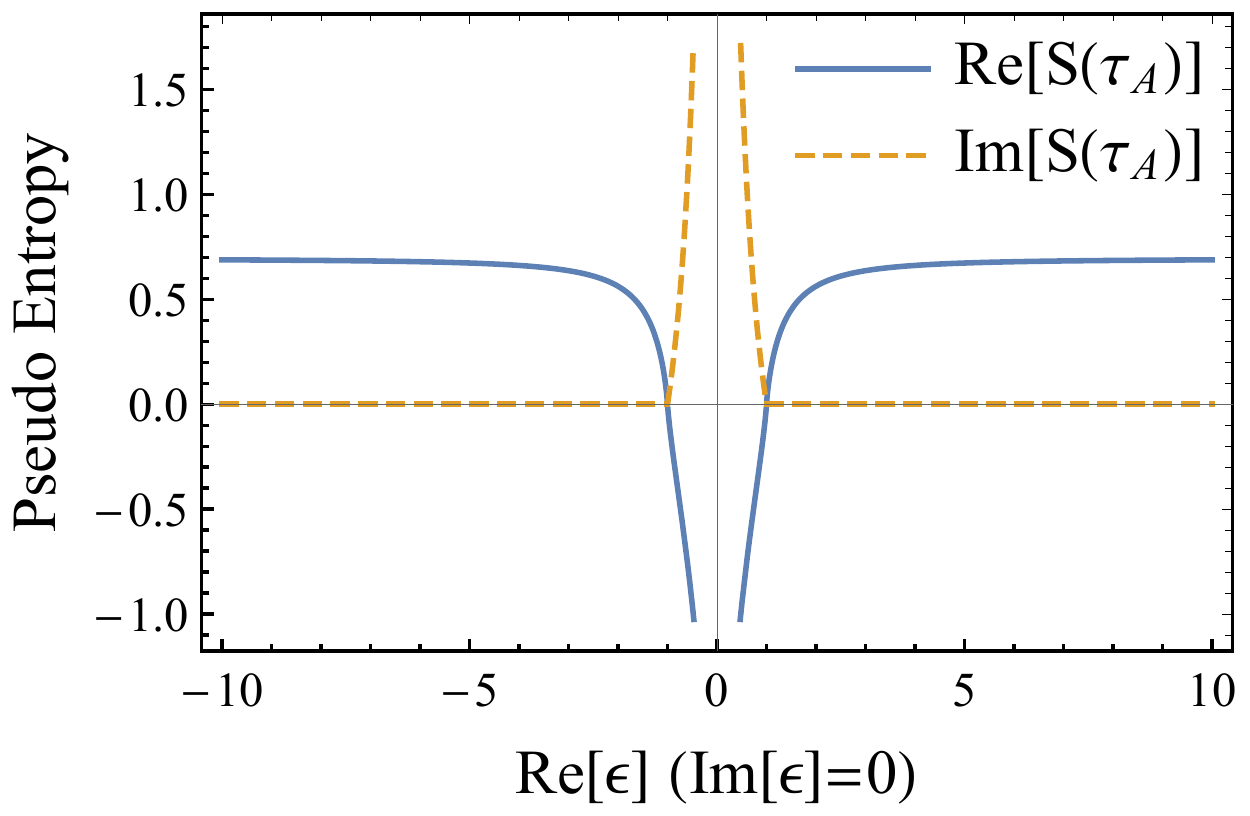}}}\subfigure[$\textrm{Re}(\epsilon)=0$]{\resizebox{80mm}{!}{\includegraphics{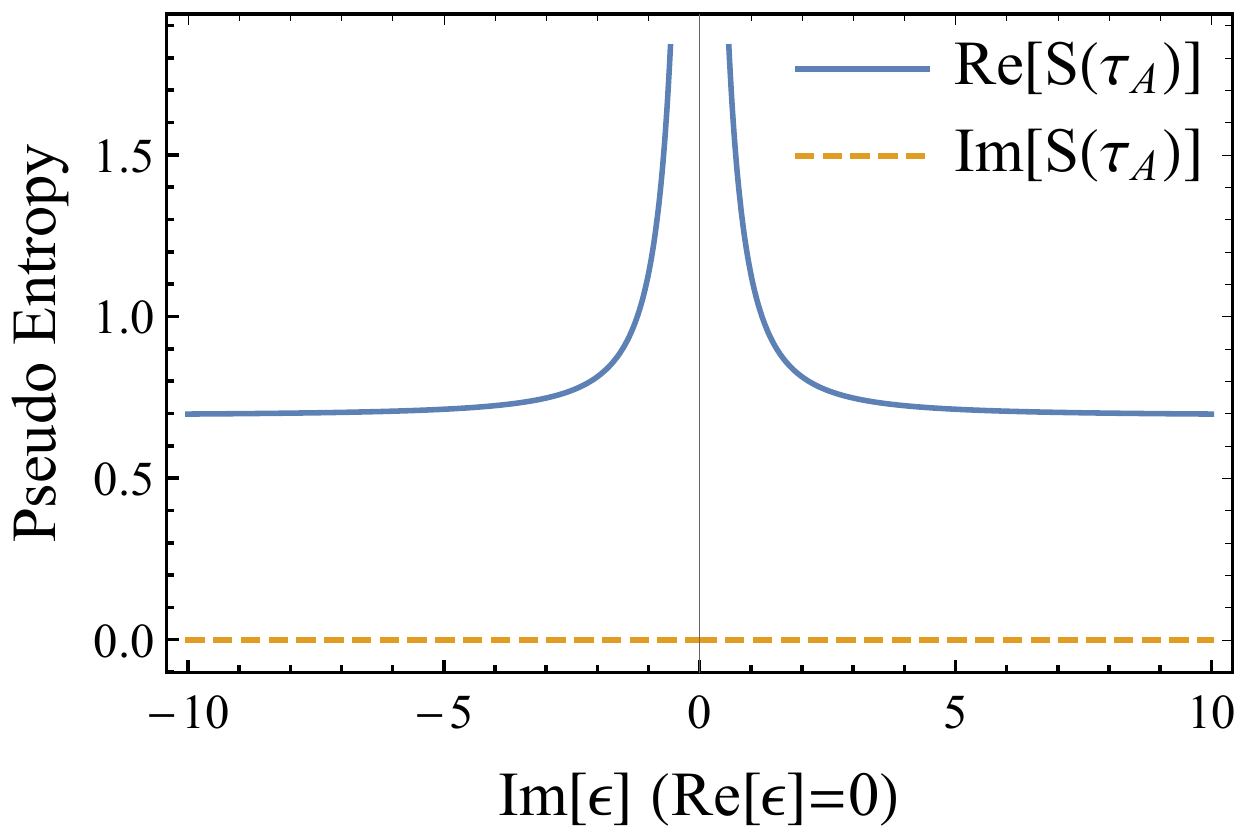}}}
    \caption{Amplification of pseudo entropy for the states \eqref{eq:psi} and \eqref{eq:varphi}. Here we show the case $\epsilon$ is (a) real or (b) pure imaginary. In the case (b), the real-valuedness comes from a pair of eigenvalues that are complex conjugates of each other. It is useful to note that the lower or upper bound of the real part is given by $\log2$ as $|\epsilon|\rightarrow\infty$ limit corresponds to the limit $\psi\rightarrow\varphi$. }
    \label{fig:qubit1}
\end{figure}

The transition matrix then has the following form,
\begin{align}
    \tau^{\psi|\varphi}_A=\dfrac{1+\epsilon}{2\epsilon}|0\rangle_A\langle0|-\dfrac{1-\epsilon}{2\epsilon}|1\rangle_A\langle1|.\label{eq:transition}
\end{align}
We stress again that we cannot define the transition matrix at the exact $\epsilon=0$ point. Nevertheless, we can imagine small but nonzero $\epsilon$ so that the resulting weak value can be amplified by the parameter $\epsilon^{-1}$. 
The pseudo entropy gives 
\begin{align}
    S(\tau^{\psi|\varphi}_A)=-\dfrac{1+\epsilon}{2\epsilon}\log\left(\dfrac{1+\epsilon}{2\epsilon}\right)-\dfrac{\epsilon-1}{2\epsilon}\log\left(\dfrac{\epsilon-1}{2\epsilon}\right).
\end{align}
Obviously, the pseudo entropy is amplified as we make $\epsilon$ smaller. See Figure \ref{fig:qubit1}. 

As is the entanglement entropy, we can also define the pseudo R\'{e}nyi entropy,
\begin{align}
    S^{(n)}(\tau^{\psi|\varphi}_A)=\dfrac{\log\mathrm{Tr}_{A}\,(\tau^{\psi|\varphi}_A)^n}{1-n}. \label{eq:pre}
\end{align}
In general, the transition matrix is no longer Hermitian, so not guaranteed to be diagonalizable. Even in such cases, we can still define trace of such matrices. Thus, strictly speaking, we have to first define the R\'{e}nyi version \eqref{eq:pre} and then define the pseudo entropy as $n\rightarrow1$ limit. 

\begin{figure}[t]
    \centering
    \subfigure[$\textrm{Im}(\epsilon)=0$]{\resizebox{80mm}{!}{\includegraphics{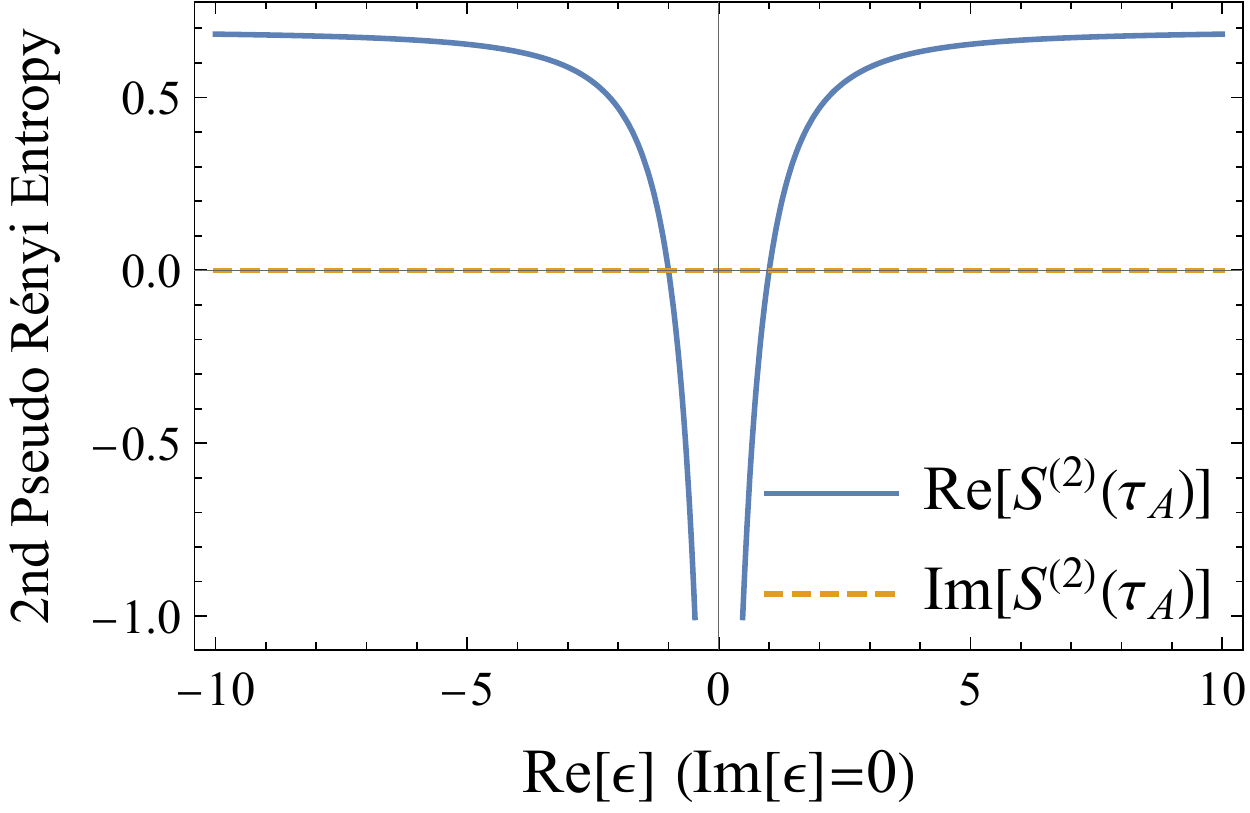}}}\subfigure[$\textrm{Re}(\epsilon)=0$]{\resizebox{80mm}{!}{\includegraphics{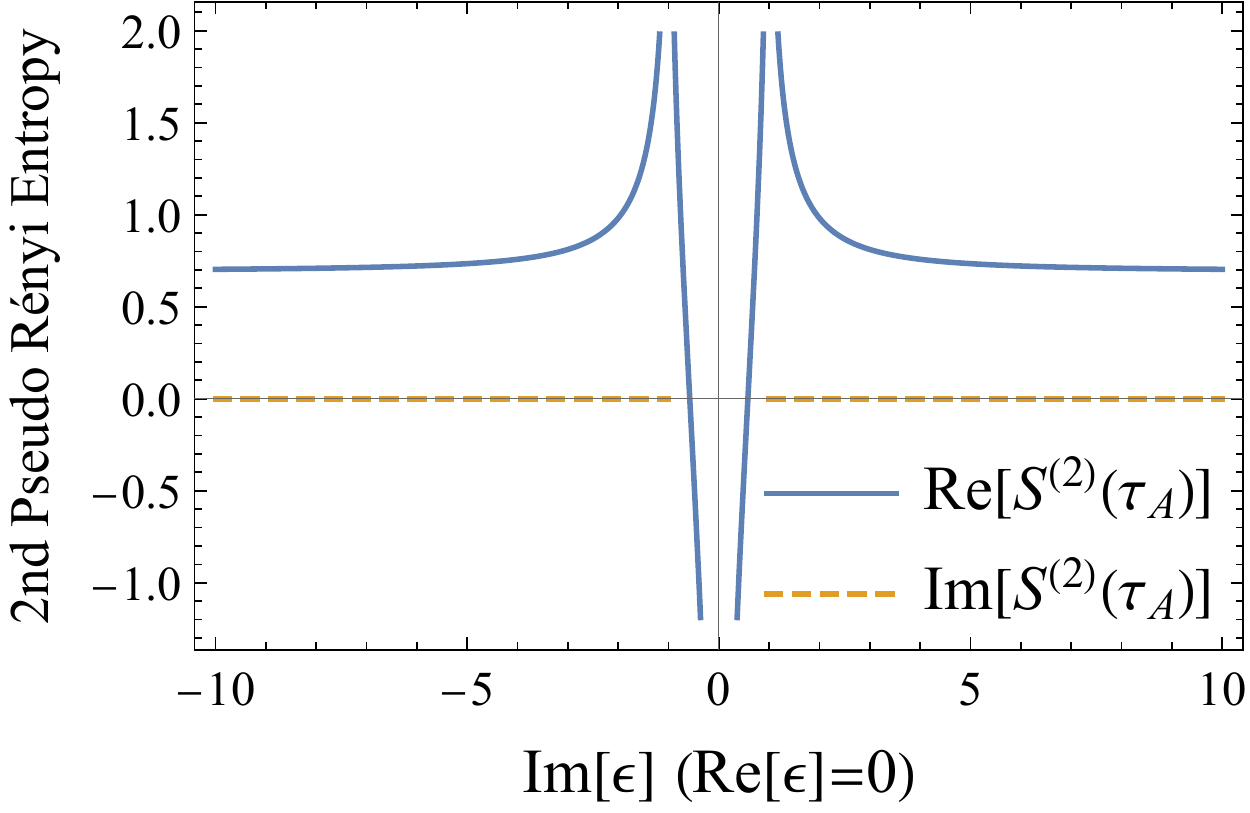}}}
    \caption{Amplification of the second pseudo R\'{e}nyi entropy for the states \eqref{eq:psi} and \eqref{eq:varphi}. Again we show the case $\epsilon$ is (a) real or (b) pure imaginary. In the case (b) with $|\epsilon|<1$, we have $\mathrm{Im}S^{(2)}=\pi$. }
    \label{fig:qubit2}
\end{figure}

For our later convenience, we note the second R\'{e}nyi result,
\begin{align}
S^{(2)}(\tau^{\psi|\varphi}_A)=\log\left(\dfrac{2\epsilon^2}{1+\epsilon^2}\right). 
\end{align}
Notice that detailed property of the amplification is quite different for each value of $n$ (See Figure \ref{fig:qubit2}). In this paper, we will mainly focus on whether pseudo entropy is amplified or not. In the current example, we see large amplifications of magnitude of $\mathrm{Re}[S^{(n)}]$ for any positive $n$. 

\subsection*{Amplification and weak value}
We would like to stress that two entangled states with small overlap does not necessarily imply the amplification. 
This is similar to the amplification of weak value,
\begin{align}
    \mathrm{Tr}(\tau^{\psi|\varphi}\hat{A})=\dfrac{\langle\psi|\hat{A}|\varphi\rangle}{\langle\psi|\varphi\rangle}. \label{eq:wv}
\end{align}
If neither initial nor finial states are eigenstates of the observable $\hat{A}$, one can amplify the weak value. Let us see examples as an illustration. Let $X$ be a spin flip operator, $X|0\rangle=|1\rangle, X|1\rangle=|0\rangle$ and $Z$ be a phase operator, $Z|0\rangle=|0\rangle, Z|1\rangle=-|1\rangle$. If we choose $\hat{A}=X\otimes X$, the weak value \eqref{eq:wv} amplifies with respect to $\epsilon$. On the other hand, if $\hat{A}=Z\otimes Z$, the weak value does not amplify since both of two states are eigenstates of $Z\otimes Z$. 

The pseudo entropy indeed shares the similar property. In the previous example, two states are not left/right eigenstates of the reduced transition matrix (or ``modular Hamiltonian'') thereof, 
\begin{align}
    (\tau_A^{\psi|\varphi})^\dagger|\varphi\rangle\neq\alpha_\varphi|\varphi\rangle,\;\tau_A^{\psi|\varphi}|\psi\rangle\neq\alpha_\psi|\psi\rangle. 
\end{align}
For the sake of comparison, let us consider a three qubit example,
\begin{align}
    \ket{\psi}&=\ket{\Phi^+}\ket{0}+\epsilon\ket{\Phi^+}\ket{1},\\
    \ket{\varphi}&=\ket{\Phi^+}\ket{1}.
\end{align}
In this case, the orthogonality stems from a decoupled sector, not relevant to the quantum entanglement, ({\it e.g.} super selection sector).
One can easily check that
\begin{align}
    \tau_{AC}^{\psi|\varphi}=\dfrac{I_A}{2}\otimes\left(\dfrac{1}{\epsilon}\ket{0}_C\bra{1}+\ket{1}_C\bra{1}\right)
\end{align}
and
\begin{align}
(\tau_{AC}^{\psi|\varphi})^\dagger|\varphi\rangle=\dfrac{1}{2}|\varphi\rangle, \;
\tau_{AC}^{\psi|\varphi}|\psi\rangle=\dfrac{1}{2}|\psi\rangle.
\end{align}
In this example, indeed we do not see the amplification, 
\begin{align}
S^{(n)}(\tau_{AC}^{\psi|\varphi})=S^{(n)}(\tau_{A}^{\psi|\varphi})=\log2.
\end{align}
As we will see later, the holographic pseudo entropy has rather a similar structure to the latter example.
\section{Amplification in 2D Free CFT}
In this section, we move to quantum field theories and discuss if the similar amplification can be realized.  
As the simplest example, we consider a two-dimensional massless free scalar $\phi$ on $\mathbb{R}^{1,1}$. We are interested in excited states which are given by primary operators acting on the vacuum state $\ket{0}$ in the 2D CFT,
\begin{align}
    \ket{\Phi^{\pm}}\equiv\mathcal{O}_\pm\ket{0} =(\mathrm{e}^{i\frac{\phi}{2}}\pm\mathrm{e}^{-i\frac{\phi}{2}})\ket{0},
\end{align}
where primary operators $\mathcal{O}_{\pm}$ have a common conformal dimension $h_{\mathcal{O}_\pm}=\bar{h}_{\mathcal{O}_\pm}=\frac{1}{8}$. 
It is well-known that the increase of (R\'{e}nyi) entanglement entropy due to quenches with these operators can be explained by EPR pairs propagating at the speed of light~\cite{Nozaki:2014hna,Nozaki:2014uaa,Caputa:2014vaa,He:2014mwa}, often referred as quasiparticle picture. For this, it is instructive to decompose the scalar field $\phi$ into left and right moving part $\phi=\phi_L+\phi_R$ and then identify $\mathrm{e}^{\pm i\frac{\phi_L}{2}}\ket{0}\sim\ket{\pm}_{L}$ and $\mathrm{e}^{\pm i\frac{\phi_R}{2}}\ket{0}\sim\ket{\pm}_{R}$. 
It means that we can treat these excited states as a QFT analog of EPR pairs
$\ket{\Phi^\pm}\sim\ket{+_L+_R}\pm\ket{-_L-_R}$ that should correspond to the states discussed in the previous section. Therefore, we again study the similar states as \eqref{eq:psi} and \eqref{eq:varphi},
\begin{align}
    \ket{\psi}&=\ket{\Phi^{+}}+\epsilon\ket{\Phi^{-}}\equiv\mathcal{O}_1 \ket{0},\\
  \ket{\varphi}&=\ket{\Phi^{-}}\equiv\mathcal{O}_2 \ket{0}.
\end{align}

\subsection*{Replica trick for pseudo R\'{e}nyi entropy}\label{sec:free}
\begin{figure}[t]
    \centering
\resizebox{100mm}{!}{\includegraphics{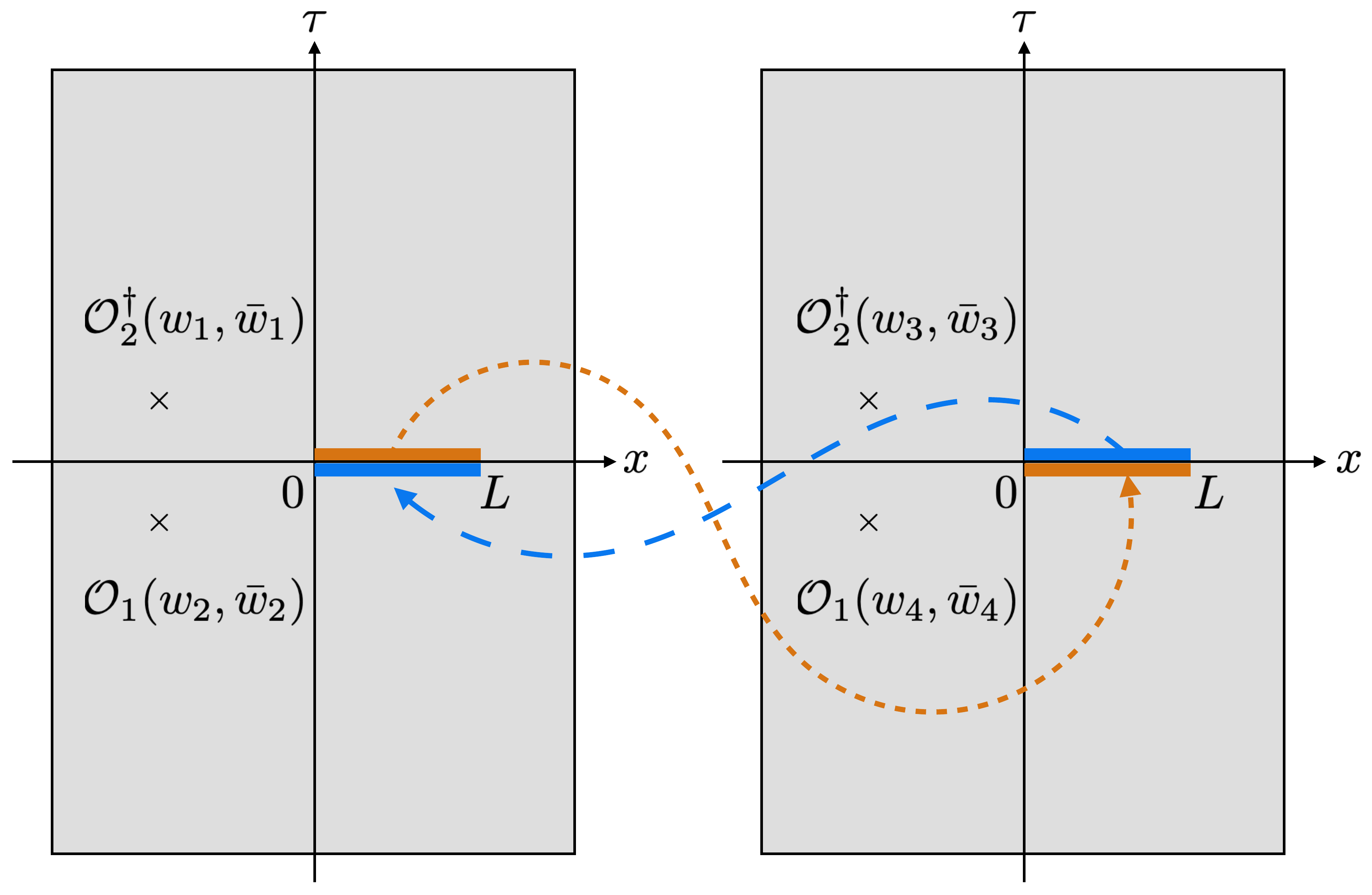}}
    \caption{Euclidean path integral on $\Sigma_2$ for calculating the numerator of the correlation function in \eqref{eq:4pt}.}
    \label{fig:sigma2}
\end{figure}

In order for computing the increase of the pseudo entropy, we first describe the replica trick for the pseudo R\'{e}nyi entropy. We consider Euclidean path integral in order for describing the $n$-th moments of transition matrices. We use complex coordinates
\begin{align}
    w=x+i\tau,\; \bar{w}=x-i\tau.
\end{align}
We insert the primary operators into $x=-\ell$ at initial time $t=0$ and take subsystem $A$ as a connected line $[0,L]$ on a fixed time slice.  
Our transition matrix is expressed as
\begin{align}
    \tau^{\psi|\varphi}(t)&=\mathcal{N}e^{-iHt}e^{-a H}\mathcal{O}_1(-\ell)\ket{0}\bra{0}\mathcal{O}_2^\dagger(-\ell)e^{-a H}e^{iHt} \nonumber \\
    &=\mathcal{N}\mathcal{O}_1(w_2,\bar{w_2})\ket{0}\bra{0}\mathcal{O}_2^\dagger(w_1,\bar{w_1}),
\end{align}
where $\mathcal{N}$ is normalization constant determined by $\mathrm{Tr}(\tau^{\psi|\varphi})=1$ and 
\begin{align}
     w_1&=-\ell+i(a-it),\;\bar{w}_1=-\ell-i(a-it),\\
     w_2&=-\ell-i(a+it),\;\bar{w}_2=-\ell+i(a+it).
\end{align}
Here $a$ is introduced to make the state normalizable and will be taken to zero later.
Note that we will treat $a\pm it$ as real numbers until the end of the calculation. 
We would like to compute the increase of the pseudo R\'{e}nyi entropy by subtracting the vacuum entanglement entropy. It is given by
\begin{align}
    \Delta S^{(n)}(\tau_{A}^{\psi|\varphi})=S^{(n)}(\tau_{A}^{\psi|\varphi})-S^{(n)}(\rho^{(0)}_{A})=\dfrac{1}{1-n}\log\left[\dfrac{\mathrm{Tr}(\tau_{A}^{\psi|\varphi})^n}{\mathrm{Tr}(\rho^{(0)}_{A})^n}\right],
\end{align}
where $\rho^{(0)}_{A}$ is the reduced density matrix for the CFT vacuum state, $\rho_A=\textrm{Tr}_B|0\rangle\langle0|$. 
One can compute $\Delta S^{(n)}$ from the $2n$-point function of the inserted operators on the $n$-sheet Riemann surface $\Sigma_n$. See \cite{Nakata:2020luh} for more detailed treatment.
We will only discuss $n=2$ case as it is enough for our purpose. We have
\begin{equation}
  \Delta S^{(2)}(\tau_{A}^{\psi|\varphi}) = -\log\left[\frac{\braket{\mathcal{O}_2^\dagger(w_1,\bar{w}_1)\mathcal{O}_1(w_2,\bar{w}_2)\mathcal{O}_2^\dagger(w_3,\bar{w}_3)\mathcal{O}_1(w_4,\bar{w}_4)}_{\Sigma_2}}{(\braket{\mathcal{O}_2^\dagger(w_1,\bar{w}_1)\mathcal{O}_1(w_2,\bar{w}_2)}_{\Sigma_1})^2}\right], \label{eq:4pt}
\end{equation}
where $\braket{\cdots}_{\Sigma_2}$ is an expectation value on the two-sheet Riemann surface $\Sigma_2$ depicted in Figure \ref{fig:sigma2}. On the other hand, $\Sigma_1$ is the original Euclidean manifold $\mathbb{R}^2$. 

For further computation of the correlation function \eqref{eq:4pt}, we consider a conformal transformation from $\Sigma_2$ to $\Sigma_1$,
\begin{align}
  \frac{w}{w-L}=z^2.
\end{align}
In particular, each point $w_i\in\Sigma_2 (i=1,\cdots,4)$ maps to $z_i\in\Sigma_1 (i=1,\cdots,4)$ as follows
\begin{align}
  z_1&=-z_3=\sqrt{(l-t-ia)/(l+L-t-ia)}, \\
  z_2&=-z_4=\sqrt{(l-t+ia)/(l+L-t+ia)}.
\end{align}
\subsection*{Amplification of pseudo R\'{e}nyi entropy}
After straightforward calculation, we obtain
\begin{equation}
  \frac{\braket{\mathcal{O}_2^\dagger(w_1,\bar{w}_1)\mathcal{O}_1(w_2,\bar{w}_2)\mathcal{O}_2^\dagger(w_3,\bar{w}_3)\mathcal{O}_1(w_4,\bar{w}_4)}_{\Sigma_2}}{(\braket{\mathcal{O}_2^\dagger(w_1,\bar{w}_1)\mathcal{O}_1(w_2,\bar{w}_2)}_{\Sigma_1})^2} = \frac{1}{2\epsilon^2}\left[(1+\epsilon^2)-(1-\epsilon^2)(|1-z|+|z|)\right],
\end{equation}
where we introduced the cross ratio 
\begin{align}
    z=\dfrac{(z_1-z_2)(z_3-z_4)}{(z_1-z_3)(z_2-z_4)}.
\end{align}
Under the time evolution with small $a$ limit, we have
\begin{align}
    (z,\bar{z})\rightarrow    \begin{cases}(0,0) & (\text{$0<t<\ell$ or $\ell+L<t$}),\\
    (1,0)& \qquad(\ell<t<\ell+L). \end{cases}
\end{align}
Therefore, we obtain
\begin{align}
        \Delta S^{(2)}(\tau_{A}^{\varphi|\psi}) = 
    \begin{cases}
        0 & (\text{$0<t<\ell$ or $\ell+L<t$}),\\
        \log\left(\frac{2\epsilon^2}{1+\epsilon^2}\right) & \qquad(\ell<t<\ell+L).
    \end{cases}
\end{align}
For $\ell<t<\ell+L$, one of the propagating EPR pairs (the right-moving particle) is in subsystem $A$, hence we obtain a non-trivial pseudo R\'{e}nyi entropy. In particular, this exactly matches the previous result in section \ref{sec:qubit}. 

\section{Non-amplification in Holographic CFT}\label{sec:hcft}
Finally, we move on to the two-dimensional holographic CFTs, dual to three-dimensional Einstein gravity in AdS. We would like to consider a similar setup as the previous sections and see if the amplification occurs. For this, we discuss a specific linear combination of the so-called heavy states, microstates of the BTZ black hole~\cite{Banados:1992wn},
\begin{align}
    \ket{\psi}&=\ket{\mathcal{O}_{H_1}}+\epsilon\ket{\mathcal{O}_{H_2}}, \label{eq:btz1}\\
    \ket{\varphi}&=\ket{\mathcal{O}_{H_2}} \label{eq:btz2},
\end{align}
where $\ket{\mathcal{O}_{H_1}}$ and $\ket{\mathcal{O}_{H_2}}$ are assumed to have different conformal dimensions $h_{H_1}$ and $h_{H_2}$. For simplicity, we consider the static black hole ($h_{H_i}=\bar{h}_{H_i}$). For being the heavy states, we assume $h_{H_i}/c\geq \frac{1}{24}$, where $c$ is the central charge. Since we assumed these heavy states have different conformal dimensions, they are orthogonal to each other,
\begin{align}
    \braket{\mathcal{O}_{H_1}|\mathcal{O}_{H_2}}=0. 
\end{align}
The conformal dimension $h$ and the radius of BTZ black hole $r_+$ are related as
\begin{align}
    r_+(h)=\sqrt{\frac{24}{c}h-1}.
\end{align}
The entanglement entropy of these heavy states $\rho^{(i)}=|\mathcal{O}_{H_i}\rangle\langle\mathcal{O}_{H_i}|$ with a small subsystem $A$ is computed as
\begin{align}
    S(\rho_A^{(i)})=\dfrac{c}{3}\log\left[\dfrac{2}{r_+(h_{H_i})\varepsilon_{\text{UV}}}\sinh\left(\dfrac{r_+(h_{H_i}) \ell_A}{2}\right)\right] \;\;\;(i=1,2), \label{eq:ee_btz}
\end{align}
where $\varepsilon_{\text{UV}}$ is the UV cutoff and $\ell_A$ is the size of subsystem $A$.  The central charge $c$ is given by the standard dictionary of AdS/CFT as $c=\frac{3}{2G_N}$, where $G_N$ is the three-dimensional Newton constant.
One can confirm the above expression from both CFT~\cite{Asplund:2014coa}~\footnote{See also \cite{Kusuki:2019rbk,Kusuki:2019evw} where the original argument~\cite{Asplund:2014coa} has been refined.} and gravity analysis~\cite{Ryu:2006bv}. For sufficiently large black holes $r_+\gg1$, the entanglement entropy \eqref{eq:ee_btz} enjoys the volume law,
\begin{align}
    S(\rho_A^{(i)})\sim\dfrac{c}{3}\dfrac{\pi \ell_A}{\beta_{H_i}},
\end{align}
where $\beta_{H_i}=2\pi/r_+(h_{H_i})$ is the inverse temperature of BTZ black hole. 
Therefore, these states are quite similar to the previous Bell states: both states are highly entangled, but orthogonal to each other. 

Let us compute the pseudo entropy for \eqref{eq:btz1} and \eqref{eq:btz2}.
The holographic pseudo entropy is computed as
\begin{align}
    S(\tau^{\psi|\varphi}_A)=\dfrac{c}{3}\log\left[\dfrac{2}{r_+(h_{H_2})\varepsilon_{\text{UV}}}\sinh\left(\dfrac{r_+(h_{H_2}) \ell_A}{2}\right)\right],\label{eqref:lin}
\end{align}
because the inner product of two states reduces to the one of the second heavy state,
\begin{align}
\braket{\varphi|\psi}=\epsilon\braket{\mathcal{O}_{H_2}|\mathcal{O}_{H_2}}.
\end{align}
Notice that we have no $\epsilon$-parameter in the above expression \eqref{eqref:lin}. 
This is also consistent with one of the expected property of holographic pseudo entropy as the weak value of area operator~\cite{Nakata:2020luh},
\begin{align}
    S(\tau^{\psi|\varphi}_A)=\dfrac{\braket{\varphi|\frac{\hat{A}}{4G_N}|\psi}}{\braket{\varphi|\psi}}.
\end{align}
One can also confirm the above results from CFT analysis as \cite{Nakata:2020luh}. Thus, we conclude that there is no amplification of pseudo entropy in this holographic example. 

However, we have to stress that the discussion so far is true only when we can regard entanglement entropy/pseudo entropy as the expectation value/weak value of the area operator. This should not be always true because the modular Hamiltonian is state-dependent and non-linear with respect to the states. In particular, for non-amplification, heavy states should be eigenstates of the area operator or reduced density/transition matrix. 

Saying differently, we expect that the area operator also has exponentially small off-diagonal elements as discussed in \cite{Almheiri:2016blp}. A schematic understanding is as follows. If we assume the eigenstate thermalization hypothesis (ETH)~\cite{Srednicki_1999,DAlessio:2015qtq,Mondaini_2017} like relation for the area operator, we have 
\begin{align}
    \langle E_i|\hat{A}|E_j\rangle =A(E)\delta_{ij}+\mathrm{e}^{-S(E)/2}g_{\mathcal{A}}(E_i,E_j)R_{ij}, \label{eq:eth}
\end{align}
where the diagonal element $A(E)$ is the thermal expectation value, $S(E)$ is the entropy at energy $E=(E_i+E_j)/2$, \,$g_{\mathcal{A}}(E_i,E_j)$ is $\mathcal{O}(1)$ function and $R_{ij}$ is the random matrix with unit variance.
The point here is that the second (off-diagonal) term is exponentially suppressed for the entropy which is of order $S=\mathcal{O}(c)$. 
In our case, since the heavy states are the energy eigenstates of the CFT Hamiltonian, we schematically obtain
\begin{align}
    \dfrac{\langle{\varphi}|\frac{\hat{A}}{4G_N}|{\psi}\rangle}{\langle{\varphi}|{\psi}\rangle}= S(\rho^{(2)}_A)+\frac{1}{\epsilon}\mathrm{e}^{-\mathcal{O}(c)}(\textrm{off-diagonal part}). \label{eq:wvarea}
\end{align}
Therefore, if we take $\epsilon$ as small as $e^{-\mathcal{O}(c)}$, we can no longer neglect the second term \eqref{eq:wvarea}\footnote{For the direct $2$D CFT analysis, it is related to the validity range of the vacuum block approximation in the Virasoro conformal block expansion.} and we will see the amplification. 

\section{Discussion}\label{sec:con}

In this paper, we have studied pseudo entropy for the particular type of linear combination in various setups and discuss when the (non-)amplification happens. In particular, the linear combination of the heavy states in holographic CFT does not give rise to amplification as long as the overlap $\braket{\varphi|\psi}$ is much larger than $e^{-\mathcal{O}(c)}$. 

An immediate but interesting future direction is to study if there are any field theoretic setups other than the holographic CFT where the amplification does not happen. It might serve as a sharp criterion whether a given QFT is in a ``holographic regime''. 
Another obvious future task is to generalize field theory results to higher dimensions. Assuming the Lewkowycz-Maldacena type argument~\cite{Lewkowycz:2013nqa}, the generalization of our holographic results to the higher dimension is straightforward, but the full understanding of the validity regime requires non-perturbative treatment. For this, it would be instructive to utilize so-called fixed area states or random tensor networks~\cite{Akers:2018fow,Dong:2018seb,Bao:2018pvs,Bao:2019fpq,Hayden:2016cfa,Penington:2019kki,Akers:2020pmf}. 

We used a very specific linear combination to compare the results for different systems.
This linear combination is ideal but appears to be artificial for amplification. One of the most interesting situations for amplification is when two states belong to different quantum phases~\cite{Mollabashi:2020yie, Mollabashi:2021xsd, Nishioka:2021cxe}. 
In these situations, one can see
\begin{align}
\Delta S(\tau_{A}^{\psi|\varphi})\equiv \dfrac{1}{2}\left[S(\tau_{A}^{\psi|\varphi})+S(\tau_{A}^{\psi|\varphi})^\ast-S(\tau_{A}^{\psi|\psi})-S(\tau_{A}^{\varphi|\varphi})\right]>0,
\end{align}
which implies the amplification discussed in this paper. 
It would be intriguing to study our setup from the viewpoint of the energy eigenstates of gapped Hamiltonians. It is also interesting to construct explicit gravity solutions that show $\Delta S(\tau_{A}^{\psi|\varphi})>0$.

Since the R\'{e}nyi entropy can be observed in the lab~\cite{Islam:2015mom}, it is a very important future problem to design measurement for the pseudo R\'{e}nyi entropy in the real experiment. We leave these interesting directions for further work.

\section*{Acknowledgement}
We are grateful to Keiju Murata, Tadashi Takayanagi, and Daisuke Yamamoto for discussion and useful comments.  
KT would like to thank the organizers and audience of ``Tensor Network States: Algorithms and Applications (TNSAA) 2021-2022 Online Workshop'' where a part of this work has been presented. 
KT is supported by JSPS Grant-in-Aid for Early-Career Scientists 21K13920 and MEXT KAKENHI Grant-in-Aid for Transformative Research Areas A ``Extreme Universe'' 22H05265.

\if(
\appendix
\section{Intermediate step}
\begin{align} 
  &\braket{\mathcal{O}_2^\dagger(w_1,\bar{w}_1)\mathcal{O}_1(w_2,\bar{w}_2)\mathcal{O}_2^\dagger(w_3,\bar{w}_3)\mathcal{O}_1(w_4,\bar{w}_4)}_{\Sigma_2}\nonumber\\
  &=\prod^{4}_{i=1} \left|\dfrac{dw_i}{dz_i}\right|^{-\frac{1}{4}}\braket{\mathcal{O}_2^\dagger(z_1,\bar{z}_1)\mathcal{O}_1(z_2,\bar{z}_2)\mathcal{O}_2^\dagger(z_3,\bar{z}_3)\mathcal{O}_1(z_4,\bar{z}_4)}_{\Sigma_1}\\
  &=L^{-1}\left|\frac{({z_1}^2-1)({z_2}^2-1)}{\sqrt{4z_1z_2}}\right| \braket{\mathcal{O}_2^\dagger(z_1,\bar{z}_1)\mathcal{O}_1(z_2,\bar{z}_2)\mathcal{O}_2^\dagger(z_3,\bar{z}_3)\mathcal{O}_1(z_4,\bar{z}_4)}_{\Sigma_1}.
\end{align}

\begin{align}
  &\braket{\mathcal{O}_2^\dagger(w_1,\bar{w}_1)\mathcal{O}_1(w_2,\bar{w}_2)}_{\Sigma_1}=2\epsilon|w_1-w_2|^{-\frac{1}{2}}=2\epsilon \left|\frac{L({z_1}^2-{z_2}^2)}{({z_1}^2-1)({z_2}^2-1)}\right|^{-\frac{1}{2}}
\end{align}

Then, the four point function can be calculated as
\begin{align} \nonumber
  &\braket{\mathcal{O}_2^\dagger(z_1,\bar{z}_1)\mathcal{O}_1(z_2,\bar{z}_2)\mathcal{O}_2^\dagger(z_3,\bar{z}_3)\mathcal{O}_1(z_4,\bar{z}_4)}_{\Sigma_1}\\
 =&4(1+\epsilon^2)\frac{\sqrt{|z_1z_2|}}{|z_1-z_2||z_1+z_2|}-(1-\epsilon^2)\frac{1}{\sqrt{|z_1z_2|}}\frac{|z_1-z_2|}{|z_1+z_2|}-(1-\epsilon^2)\frac{1}{\sqrt{|z_1z_2|}}\frac{|z_1+z_2|}{|z_1-z_2|}.
\end{align}
By using these results, we obtain
\begin{align} \nonumber
  &\frac{\braket{\mathcal{O}_2^\dagger(w_1,\bar{w}_1)\mathcal{O}_1(w_2,\bar{w}_2)\mathcal{O}_2^\dagger(w_3,\bar{w}_3)\mathcal{O}_1(w_4,\bar{w}_4)}_{\Sigma_2}}{(\braket{\mathcal{O}_2^\dagger(w_1,\bar{w}_1)\mathcal{O}_1(w_2,\bar{w}_2)}_{\Sigma_1})^2} \\ \nonumber
  =&\frac{1}{4\epsilon^2}\left(-2(1-\epsilon^2)\left|\frac{(z_1+z_2)^2}{4z_1z_2}\right|+2(1+\epsilon^2)-2(1-\epsilon^2)\left|\frac{(z_1-z_2)^2}{4z_1z_2}\right|\right).
\end{align}
)\fi
\bibliographystyle{JHEP}
\bibliography{ref.bib}
\end{document}